\documentclass[twocolumn,english,aps,pra,reprint, superscriptaddress,showpacs,longbibliography,showkeys]{revtex4-2}
\usepackage{amsmath,amssymb,bbm,mathrsfs,bm,braket,color,graphicx,comment,xcolor, colortbl,dsfont,multirow, adjustbox, booktabs}
\usepackage[colorlinks,citecolor=blue,urlcolor=blue,linkcolor = blue]{hyperref}
\usepackage[mathscr]{euscript}
\usepackage{tikz}
\usetikzlibrary{quantikz2}
\usetikzlibrary{arrows.meta}
\usepackage[normalem]{ulem}

\newcommand{\calC}{\mathcal{C}}
\DeclareMathOperator{\Tr}{Tr}
\newcommand{\bfq}{{\boldsymbol{q}}}
\newcommand{\bfbeta}{{\boldsymbol{\beta}}}
\newcommand{\bfgamma}{\boldsymbol{\gamma}}
\newcommand{\bfOmega}{\boldsymbol{\Omega}}
\newcommand{\eeuler}{\mathrm{e}}

\newcommand{\rmmin}{\mathrm{min}}
\newcommand{\rmmax}{\mathrm{max}}
\newcommand{\squaredots}{
    \vspace{-.175em}
    \tikz[line cap=round, line join=round]{
    \draw[black] (0.8ex,0.8ex) -- (0ex,0.8ex) -- (0ex,0ex) --  (0.8ex,0ex); 
    \draw[color=black, fill=black] (0ex,0ex) circle (0.175ex);
    \draw[color=black, fill=black] (0ex,0.8ex) circle (0.175ex);
    \draw[color=black, fill=black] (0.8ex,0ex) circle (0.175ex);
    \draw[color=black, fill=black] (0.8ex,0.8ex) circle (0.175ex);
    }
}

\newcommand{\parity}{Parity}
\newcommand{\Parity}{Parity}

\begin{document}

 \title{Higher-order Parity Constraints for Quantum Approximate Optimization}

\author{Anita Weidinger}
\affiliation{Institut für Theoretische Physik, Universität Innsbruck, A-6020 Innsbruck, Austria}
\affiliation{Parity Quantum Computing GmbH, A-6020 Innsbruck, Austria}
\author{Glen Bigan Mbeng}
\affiliation{Institut für Theoretische Physik, Universität Innsbruck, A-6020 Innsbruck, Austria}
\affiliation{Parity Quantum Computing GmbH, A-6020 Innsbruck, Austria}
\author{Roeland ter Hoeven}
\affiliation{Parity Quantum Computing GmbH, A-6020 Innsbruck, Austria}
\author{Wolfgang Lechner}
\affiliation{Institut für Theoretische Physik, Universität Innsbruck, A-6020 Innsbruck, Austria}
\affiliation{Parity Quantum Computing GmbH, A-6020 Innsbruck, Austria}
\affiliation{Parity Quantum Computing Germany GmbH, 20095 Hamburg, Germany} 

\date{\today}

\begin{abstract}
The Parity Architecture was originally introduced in the context of quantum optimization algorithms to overcome hardware connectivity constraints. Here, we show that its benefits extend beyond this role by identifying specific parity transformations that enhance the success probability of the Quantum Approximate Optimization Algorithm (QAOA) for optimization problems defined on hardware-native and sparse graphs. Numerical simulations suggest that this enhancement originates from the higher-order operators introduced by the Parity representation, which accelerate the propagation of correlations across the physical device.
We investigate this mechanism for the Signed Max-Cut problem on 3-regular graph instances and demonstrate its practical relevance on IBM Heron quantum processors. QAOA parameters are first classically optimized on small training instances containing up to 14 vertices and subsequently transferred to larger instances executed on quantum hardware. We consider Signed Max-Cut instances of up to 40 vertices and find that, together with an error mitigation technique, the parity-based QAOA identifies the optimal solutions for all the considered 40-vertex instances, whereas the conventional QAOA fails under the same conditions.
\end{abstract}
\maketitle

\section{Introduction}
The Quantum Approximate Optimization Algorithm (QAOA) is a hybrid quantum-classical algorithm that has emerged as a promising strategy to solve combinatorial optimization problems on near-term quantum hardware~\cite{farhi_arXiv2014, Blekos_PhysicsReports2024}. 
Its flexible variational framework and potential for shallow circuit implementations have motivated extensive theoretical and experimental investigations~\cite{Zhou_Lukin_prx2020, Pagano_PNAS2020,Harrigan_NaturePhysics2021,He_NatureComPhys2025,Boulebnane_PRL2026}.
Nevertheless, achieving a quantum advantage with QAOA remains an open challenge.

Recent results highlight QAOA's potential on selected problem classes. Studies of Max-Cut on 3-regular graphs with large girth~\cite{farhi_arXiv2026}, the random $8$-SAT problem~\cite{Boulebnane_PRXQuantum2024} and the low autocorrelation binary sequences (LABS) problem~\cite{Shaydulin_SciAdv2024} have provided evidence of favorable scaling relative to leading classical algorithms on selected optimization problems. However, the circuit depths needed to realize these potential advantages pose substantial challenges for current quantum hardware.  At small layer counts, the limited propagation of
correlations across the problem graph can restrict the algorithm's performance~\cite{farhi_arXiv2020, Bravyi_Tang_PRL2020}.

Several approaches have been proposed to improve QAOA's performance at feasible circuit depths~\cite{Blekos_PhysicsReports2024}, including recursive methods~\cite{Bravyi_Tang_PRL2020, Bae_Lee_Springer2024}, warm starts~\cite{Egger_quantum2021, Tate_Gupta_Quantum2023, Yunglong_arXiv2025}, more expressive circuits~\cite{Herrman_ScientificReports2022, Gaidai_scientificReports2024, Vijendran_IOPscience2024} and adaptive choices of mixer
operators~\cite{Yunlong_PRR2022, Zhu_Economou_PRR2022}. Increasing the number of layers also increases the number of parameters to optimize and exposes the circuit to more noise. 
Approaches based on parameter concentration and transferability were introduced to reduce the classical optimization effort~\cite{Wurtz_PRA2021,Galda_IEEE2021,Guo_Egger_arXiv2026}, whereas error mitigation~\cite{Li_Benjamin_PRX2017,Temme_PRL2017,Endo_Li_PRX2018,McClean_PRA2017,Weidinger_PRA2023}, error detection~\cite{He_CommPhys2025}, and quantum error correction~\cite{Perlin_arxiv2026} strategies were proposed to improve the reliability of deeper QAOA circuits. 

Hardware connectivity presents a further obstacle for QAOA. When the edges of the problem graph do not match the connections between qubits on the processor, a direct implementation of the QAOA circuit generally requires routing operations~\cite{Harrigan_NaturePhysics2021}. These operations increase the circuit's gate count, depth, and exposure to noise, motivating the development of compilation methods to reduce routing overhead~\cite{tket,Berend_PRR2026,Fleckenstein_Dreier_IEEE2025}.

The Parity transformation addresses this connectivity mismatch by encoding logical interactions in physical qubits and imposing consistency through Parity constraints~\cite{Lechner_2015,Lechner_IEEE2020}.  The original Parity QAOA construction uses local three- and four-qubit constraints that can be implemented in parallel on a suitable hardware layout, allowing each QAOA layer to be realized with a shallow circuit whose depth is independent of the problem size. Moreover, the redundancy introduced by the Parity transformation can be exploited for error mitigation~\cite{Weidinger_PRA2023}. However, restricting the constraints to three- or four-qubit operators can slow the spread of correlations across the encoded problem. In these cases, Parity QAOA may require additional layers to match the performance of conventional QAOA, potentially offsetting the reduction in circuit depth per layer~\cite{Wybo_Quantum2024,Weidinger_NJP2026}.

In this work, we exploit the flexibility of the parity transformation to improve Parity QAOA's performance on sparse graphs using multi-qubit constraints involving more than four physical qubits. These constraints couple larger portions of the encoded problem within each QAOA layer and may reduce the
number of layers needed to establish long-range correlations. Their
implementation can, however, increase the circuit depth per layer. Wang et al.~\cite{Wang_Weidinger_arXiv2026} previously demonstrated the benefits of this approach for frustrated Ising rings~\cite{Knysh_PRA2020}, showing that Parity QAOA can reach optimal solutions with fewer layers than conventional QAOA.
Here, we extend the study to more complex problems on sparse graphs and investigate whether, and to what extent, the use of larger constraints
translates into improved optimization performance

In this manuscript, we first introduce the signed Max-Cut problem in Sec~\ref{sec:signed_maxcut}, QAOA and Parity QAOA in Secs.~\ref{sec:QAOA} and \ref{sec:ParityQAOA}, respectively, and the covering depth in Sec.~\ref{sec:see_graph}. We then provide results for a (hardware native) heavy-hex graph, in Sec.~\ref{sec:results:hardware_native}, which include noiseless simulations as well as hardware experiments on an IBM Heron device~\cite{IBM_quantum}. Finally, we investigate random 3-regular graphs in Sec.~\ref{sec:3-regular_graphs}, where we
compare compiled circuits for problems with up to $40$ vertices and present noiseless
simulations on smaller instances alongside hardware experiments.

\section{Background}

\subsection{The Signed Max-Cut problem}\label{sec:signed_maxcut}
Building on previous work \cite{Weidinger_NJP2026, Wybo_Quantum2024}, we consider the signed Max-Cut problem throughout this work. In general, this problem is NP-hard~\cite{Garey_JACM1978,Garey_TCS1976} and potential fields of application include social networks, systems biology, and integrated circuit design~\cite{Chiang_Zelikovsky_IEEE2007,DasGupta_Zhang_Biosystems2007,Huffner_Niedermeier_book2007,Zaslavsky_ElectronicElectronicofCombinatorics2021}. 

Given an undirected graph ${G=(V, E)}$ with vertices $V$ and edges ${E \subseteq V^2}$, each edge ${(i, j) \in E}$ is associated with a weight ${J_{ij} \in \{-1, +1\}}$, where positive weights favor placing the connected vertices in different subsets and negative weights favor placing them in the same subset.
The goal in the signed Max-Cut problem is to find a graph bipartition that maximizes the weighted sum of ``cut'' edges, i.e. 
 a partition of the vertex set $V$ into two disjoint subsets, $V_1$ and $V_2$, such that ${V_1 \cup V_2 = V}$, with the objective to maximize the sum of the weights of the edges that connect vertices across the two subsets. This problem can be described by the objective function 
\begin{equation}\label{eq:objective_function}
C(\boldsymbol{s}) = -\sum_{(i, j) \in E} J_{ij} (s_i \oplus s_j),
\end{equation}
where ${s_i = 0}$ or ${s_i = 1}$ indicates whether vertex $i$ is assigned to subset $V_1$ or $V_2$, respectively, and $\oplus$ denotes an addition modulo 2.
Any bitstring ${\boldsymbol{s} = (s_1, s_2, \dots, s_{N}) \in \{0, 1\}^N}$ (${N=|V|}$) that minimizes Eq.~\eqref{eq:objective_function} is a solution to the optimization problem.

To obtain the Hamiltonian corresponding to the classical objective function, we assign a qubit to each vertex of the graph, giving
\begin{equation}\label{eq:Hp-qubo}
H_P = -\sum_{(i, j) \in E} J_{ij} (1-\sigma_i^z \sigma_j^z),
\end{equation}
where $\sigma_i^z$ represents the Pauli-$z$ operator acting on qubit $i$. The eigenvalue of $\sigma_i^z$ corresponds to the assignment of vertex $i$ to one of the two subsets, with eigenvalues $+1$ and $-1$ indicating membership in $V_1$ or $V_2$, respectively. In the computational basis $\bra{\boldsymbol{s}}H_P \ket{\boldsymbol{s}}=2C(\boldsymbol{s})$, so the two objectives differ only by
a positive global scaling factor.
Thus, finding a ground state of $H_P$ is equivalent to finding an optimal partition that maximizes the sum of the signed edge weights between the two subsets, i.e., minimizes Eq.~\eqref{eq:objective_function}. 

\subsection{Quantum Approximate Optimization}\label{sec:QAOA}
QAOA~\cite{farhi_arXiv2014} is designed to find approximate solutions to combinatorial optimization problems, such as the signed Max-Cut problem. 
The problem is encoded in the problem Hamiltonian $H_P$, Eq.~\eqref{eq:Hp-qubo}. Starting from the initial state $\ket{\Psi_0}=\ket{+} ^ {\otimes N}$, the phase-separation unitary operator ${U_P(\gamma) = \eeuler^{-i\gamma H_P}}$  is applied alternately with the mixer unitary
${U_x(\beta) = \prod_{j=1}^{N} \eeuler^{-i\beta\sigma^x_j}}$ for $p$ layers. This results in the variational state
\begin{equation}
    \label{eq:QAOA_circuit}
    \ket{\Psi(\bm{\beta},\bm{\gamma})} =  U_x(\beta_p) U_P(\gamma_p)\dots U_x(\beta_1) U_P(\gamma_1)\ket{\Psi_0}
\end{equation}
which depends on $2p$ variational parameters ${\bm{\gamma}=(\gamma_1,\dots, \gamma_p)}$ and ${\bm{\beta}=(\beta_1,\dots, \beta_p)}$. The optimized parameters $(\bm{\beta^*}, \bm{\gamma^*})$ are obtained using a classical optimizer that seeks to minimize the objective function 
\begin{equation}
\calC(\bfbeta,\bfgamma)= \bra{\Psi(\bfbeta,\bfgamma)} H_P \ket{\Psi(\bfbeta,\bfgamma)}.
\end{equation}
The objective function is estimated by repeatedly preparing and measuring the variational state $\ket{\Psi(\bm{\beta},\bm{\gamma})}$. 

The integer $p$ denotes the number of QAOA layers. In the noiseless setting, the globally optimized
objective value approaches the ground-state energy as $p\to\infty$~\cite{farhi_arXiv2014, mbeng_fazio_santoro_arXiv2029}, while finite $p$ generally yields approximate solutions when measuring $\ket{\Psi(\bfbeta^*,\bfgamma^*)}$. 

Increasing $p$ cannot worsen the globally optimized objective value in the ideal noiseless setting. However, in practice, one limits $p$ to control the classical optimization cost and the noise accumulated during circuit
execution. The application of two-qubit gates is one of the main sources of noise in QAOA. 
For example, an $R_{zz}$ gate implements the
evolution generated by an interaction $\sigma^z_i \sigma^z_j$ in the problem Hamiltonian $H_P$, see also Fig.~\ref{fig:HeavyHex_coveringdepth_parameters} (a) and Fig.~\ref{fig:HeavyHex_coveringdepth_parameters} (c). On hardware that does not support $R_{zz}$ gates, this gate can be decomposed into two CNOT gates and a single-qubit z-rotation.

The ground-state probability (or success probability) $P_{\mathrm{GS}}$ is the probability of measuring any optimal bitstring. Let $\Pi_{\mathrm{GS}}$ denote
the projector onto the full ground-state subspace of $H_P$, then
\begin{equation}\label{eq:overlap_ground_state}
    P_{\mathrm{GS}} = \bra{\Psi(\bfbeta^*,\bfgamma^*)} \Pi_{\mathrm{GS}}\ket{\Psi(\bfbeta^*,\bfgamma^*)}.
\end{equation}

In general, the unitary $U_P({\gamma})$ cannot be implemented directly on a quantum chip if required
interactions cannot be mapped to the device connectivity. Routing is then necessary~\cite{tket, Berend_PRA2026, Berend_PRR2026, Fleckenstein_Dreier_IEEE2025},  introducing additional two-qubit gates that increase the circuit's exposure to noise.
An alternative QAOA variant that was originally designed to address this routing problem is {\parity} QAOA~\cite{Lechner_2015, Ender2023}, which is outlined in the next section.

\subsection{Parity Quantum Approximate Optimization}
\label{sec:ParityQAOA}
The Parity transformation~ \cite{Lechner_2015, Lechner_IEEE2020, Ender2023} maps the logical problem graph to a parity graph by mapping each two-qubit interaction $\sigma_i^z \sigma_j^z$ to a single-qubit term ${\tilde{\sigma}_{\nu}^z}$. This single-qubit term represents the parity of a pair of logical qubits ${\sigma_{i_{\nu}}^z\sigma_{j_{\nu}}^z \to \tilde{\sigma}_{\nu}^z}$, with one
physical qubit per logical edge, so that $K = |E|$. For the graphs considered here, $N \leq K$.

After dropping the additive constant$-\sum_{(i, j) \in E} J_{ij}$, the parity transformation maps the problem Hamiltonian of Eq.~\eqref{eq:Hp-qubo} to a sum of local field terms
\begin{equation}\label{eq:local_field_hamiltonian}
    \tilde{H}_{z}= \sum_{\nu=1}^K \tilde{J}_{\nu}\tilde{\sigma}_\nu^z
\end{equation}
where ${\tilde{J}_{\nu}=J_{i_{\nu}j_{\nu}}}$  with ${\nu=1, 2, ...K}$ are the local fields of the physical qubits. 

For a connected logical graph, $L=K-N+1$  independent constraints are needed to restrict the physical configurations to those corresponding to logical configurations, which are unique up to a global
spin flip~\cite{Ender2023}. 
These constraints can be constructed from independent cycles in the logical graph \cite{Lechner_2015, Ender2023}

The operator for the $l$-th constraint, acting on $N_{c_l}$ 
physical qubits, is
\begin{equation}\label{eq:constraint_hamiltonian}    \tilde{H}_{\squaredots,l}=\prod_{c=1}^{N_{c_l}} \tilde{\sigma}_{\nu_{l,c}}^z  \qquad \mbox{for}\;\; l=1,2, \dots,L\ ,
\end{equation}
where the integers $\nu_{l,c}$ represent the indices of the physical qubits appearing in the $l$-th parity constraint. A valid parity configuration is an eigenvector of eigenvalue $+1$ for every constraint operator.

For a complete graph, all $L$ constraints can be implemented locally in a planar square-grid geometry, the resulting triangular layout is often referred to as the Lechner-Hauke-Zoller (LHZ) model ~\cite{Lechner_2015, Lechner_IEEE2020}.
Ender et al.~\cite{Ender2023} extended the parity transformation to arbitrary graphs and hypergraphs, using algorithms that identify
suitable layouts for the physical qubits and constraint terms~\cite{parityos}. 
These implementations typically use
three- and four-body constraints, with $N_{c_l} \in \{3, 4\}$.

The problem of finding a ground state of the original Hamiltonian $H_P$ is equivalent to the problem of finding a ground state of the physical problem Hamiltonian 
\begin{equation}\label{eq:parity-qaoa_cost_function}
    \tilde{H}_P = \tilde{H}_{z} - C\sum_{l=1}^L \tilde H_{\squaredots,l}\,,
\end{equation}
where $C>0$ is a positive penalty strength. For a sufficiently large positive penalty strength~\cite{Lanthaler_NJPhys2021}, the ground states of the physical problem
Hamiltonian in Eq.~\ref{eq:parity-qaoa_cost_function}  satisfy the parity constraints and encode ground states of $H_P$. 

In {\parity} QAOA \cite{Lechner_IEEE2020,Weidinger_PRA2023}, the local-field and constraint parts of the phase-separation unitary are
assigned independent variational parameters. The corresponding unitaries are ${\tilde{U}_{z}(\gamma) = \eeuler^{-i\gamma\tilde{H}_{z}}}$ and $ {\tilde{U}_{\squaredots}(\Omega) = \prod_{l=1}^{L}\eeuler^{-i\Omega \tilde{H}_{\squaredots,l}}}$, where $\Omega$ is an additional variational parameter. This extra variational freedom
can improve performance compared with using a single phase-separation parameter, as demonstrated in Ref.~\cite{Lechner_IEEE2020}. 
Alternating these unitaries with the mixer unitary ${\tilde{U}_x(\beta) = \prod_{\nu=1}^{K} \eeuler^{-i\beta\Tilde{\sigma}_{\nu}^x}}$ $p$ times starting from the initial state ${\ket{\tilde \Psi_0}=\ket{+}^{\otimes K}}$ results in the variational state
\begin{equation}
\begin{split}
\ket{\tilde{\Psi}(\bfbeta,\bfOmega,\bfgamma)} &=  \tilde{U}_x(\beta_p) \tilde{U}_{\squaredots}(\Omega_p) \tilde{U}_z(\gamma_p)\dots\\
&\hspace{1.2cm}\dots\tilde{U}_x(\beta_1) \tilde{U}_{\squaredots}(\Omega_1) \tilde{U}_z(\gamma_1)\ket{\tilde \Psi_0}.
\end{split}
\label{eq:Parity_circuit}
\end{equation}

The Parity mapping transfers the non-locality from the problem interactions to the constraint unitaries. 
The single-qubit rotations $\tilde{U}_z(\gamma)$ and $\tilde{U}_x(\beta)$ are independent of the underlying topology, whereas the constraint unitaries $\tilde{U}_{\squaredots}(\Omega)$ must be compiled to the available connectivity. The constraints can be implemented 
with CNOT chains and $R_{zz}$ gates, see Fig.~\ref{fig:HeavyHex_coveringdepth_parameters}(b) and (c)~\cite{Lechner_IEEE2020, Weidinger_PRA2023}. 
Another advantage is that the constraints can be applied in parallel so that the algorithm can be performed at constant depth at fixed $p$~\cite{Lechner_IEEE2020, Unger_Lechner_arXiv2023}. For the LHZ triangular
layout with local three- and four-body constraints, these operations can be scheduled in a constant
number of parallel steps on a planar quantum processor, giving a circuit depth independent of $N$ at fixed $p$~\cite{Lechner_IEEE2020, Unger_Lechner_arXiv2023}.

Measuring the
physical $K$-qubit variational state produces a bitstring ${\tilde{\bfq}=(\tilde{q}_1,\dots,\tilde{q}_K)}$, which can be decoded into a configuration of the $N$ logical variables ${\bfq=(q_1,\dots,q_N)\in\{0, 1\}^N}$. In particular, only a subset of the physical qubits is necessary to decode into logical variables $\bfq$. This subset may consist of the physical qubits associated with
the parities of the $N-1$ edges of a spanning tree in the logical graph. Fixing one logical bit determines the remaining bits from these parities.
For example, knowing the parities of the pairs $(1,2)$ and $(2,3)$ in the complete graph with three nodes is sufficient to determine the logical variables $q_1$ and $q_2$, once $q_0$ is fixed. Therefore, the parity of pair $(2,3)$ is redundant for a constraint-satisfying physical configuration. 

Different spanning trees $t$, and therefore different subsets of measured parities, can yield different decoded configurations ${\bfq}^{(t)}(\tilde{\bfq})$ when the measured bitstring of the physical qubits $\tilde{\bfq}$  violates at least one parity constraint. Weidinger et al.~\cite{Weidinger_PRA2023} showed that this variability can be exploited for error mitigation. Following Refs.~\cite{Weidinger_PRA2023,Weidinger_NJP2026}, each measured bitstring $\tilde{\bfq}$ is decoded using a set of $M$ spanning trees ${\mathcal{T}=\{t_1, t_2,\dots t_M\}}$, and the logical configuration with the lowest energy is kept. We denote this configuration by $\bfq^{(\mathcal{T})}(\tilde{\bfq})$. The objective function is the average energy of these selected configurations, which reads
\begin{align}
     \tilde{\calC}(\bfbeta,\bfOmega,\bfgamma)
     = \Tr  \left( H_P \,D_{\mathcal{T}}[
     \tilde{\Psi}(\bfbeta,\bfOmega,\bfgamma)]\right)\label{eq:C_parity_min}.
\end{align}
Here $D_{\mathcal{T}}$ is the decoding map, defined as
\begin{equation}
D_{\mathcal{T}}[\tilde{\Psi}(\bfbeta,\bfOmega,\bfgamma)]=\!\!\!\sum_{\scriptscriptstyle\tilde{\bfq}\in \{0,1\}^K} \!\!\! |\braket{\tilde{\bfq}|\tilde{\Psi}(\bfbeta,\bfOmega,\bfgamma)}|^2 P_{\bfq^{(\mathcal{T})}(\tilde{\bfq})}.
\end{equation}
where $P_{\bfq^{(\mathcal{T})}(\tilde{\bfq})} = \ket{\bfq^{(\mathcal{T})}(\tilde{\bfq})}\bra{\bfq^{(\mathcal{T})}(\tilde{\bfq})}$ is the projector onto the selected logical state. The ground state probability of {\parity} QAOA's output state, after the decoding procedure, is 
\begin{equation}\label{eq:overlap_ground_state_parity}
    P_{GS} = \Tr\left(\Pi_{\mathrm{GS}}D_{\mathcal{T}}[
     \tilde{\Psi}(\bfbeta^*,\bfOmega^*,\bfgamma^*)]\right),
\end{equation}
where a measurement shot is successful if at least one of the $M$ trees decodes to a ground-state
configuration.

Following Ref.~\cite{Wang_Weidinger_arXiv2026}, in this work, we extend previous approaches by allowing constraints with $N_{c_l}>4$, which we call higher-order constraints, so that each constraint acts on a larger portion of the graph. The resulting trade-off between covering depth and gate depth per layer is discussed in Sec.~\ref{sec:see_graph}.

\subsection{Covering depth $c_p$ and higher-order constraints}\label{sec:see_graph}
The three- and four-body {\parity} constraints used in previous works can be implemented with a circuit depth per QAOA layer that is independent of system size on suitable layouts~\cite{Lechner_IEEE2020, Ender2023, Weidinger_PRA2023}. 
However, their limited support can slow the spread of correlations across the encoded problem. Additional QAOA layers may be required to match the performance of conventional QAOA, potentially offsetting the reduction in circuit depth per layer~\cite{Wybo_Quantum2024,Weidinger_NJP2026}. 
Consequently, circuit depth and gate count alone are insufficient to determine the optimization algorithm's performance. 
Following Gaidai et al.~\cite{Gaidai_scientificReports2024}, we therefore also consider the \textit{covering depth} $c_p$, which quantifies the number of QAOA layers needed for information to propagate across the entire system. 

The covering depth $c_p$ is obtained from the maximum depth of an edge-based breadth-first search (BFS)~\cite{Gaidai_scientificReports2024, farhi_arXiv2020}. 
For direct QAOA, we create an auxiliary line graph by representing each edge of the original problem graph as a vertex. Two vertices of the auxiliary graph are connected if their corresponding edges in the original graph share a vertex. 
For {\parity} QAOA, each vertex of the auxiliary graph represents a constraint in the parity Hamiltonian. In this case, vertices of the auxiliary graph are connected if the corresponding constraints share at least a physical qubit. 
The covering depth $c_p$ is, then, defined as the maximum depth of a BFS from all vertices of the auxiliary graph ~\cite{Gaidai_scientificReports2024, farhi_arXiv2020}.

In addition, we want to define the average number of qubits in the entangling gates, denoted as $C_{len}$. For the logical graph, $C_{len}=2$ for any $N$ as there are 2-qubit interactions only and for {\Parity} QAOA it is the average size of the higher-order constraint, $C_{len}=\frac{1}{L}\sum_{l=1}^L N_{c_l}$.

In this work, we focus on sparse graphs and allow higher-order constraints for {\parity} QAOA, i.e., ${N_{c_l}>4}$. In particular, we investigate how Parity QAOA with these constraint terms compares with conventional QAOA, which directly implements the logical problem Hamiltonian. We refer to the latter as \textbf{direct} QAOA and to the implementation using the parity mapping as \textbf{Parity} QAOA.

\section{Results}
We first investigate a small hardware-native graph
and then extend the analysis to random 3-regular graphs. In both cases, we compare the resulting circuits in terms of two-qubit gate depth, two-qubit
gate count, and covering depth, and assess their optimization performance in noiseless simulations and
hardware experiments.

\begin{figure*}
    \includegraphics[scale=0.36]{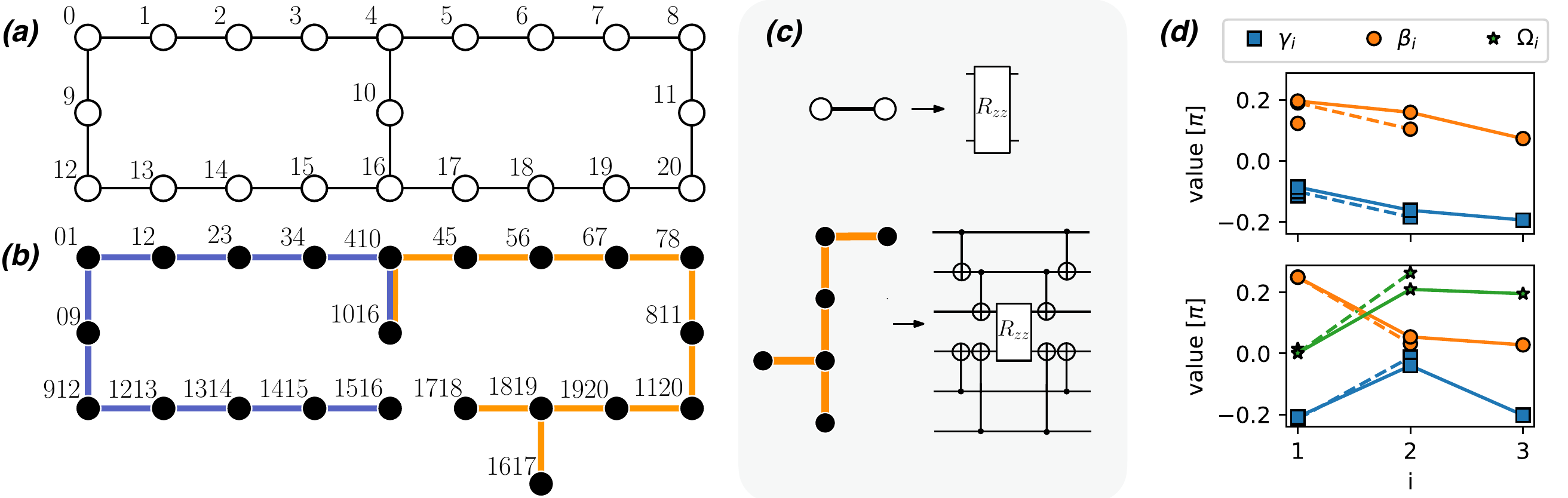}
    \caption{Heavy-hex graph with $N=21$ nodes in its logical (a) and {\Parity} (b) representation. Interactions in the logical graph can be implemented via $R_{zz}$ gates and constraints in the {\parity} graph with CNOT and $R_{zz}$ gates, shown next to the graphs in (c).
    Panel (d) shows the best found parameters for $p\in\{1, 2, 3\}$ during optimization. Parameters at $p=1$ are denoted by a single marker, while parameters at $p=2$ and $p=3$ with two or three markers that are connected via a dashed and solid line, respectively.}\label{fig:HeavyHex_coveringdepth_parameters}
\end{figure*} 

\begin{figure}
    \includegraphics[scale=0.75]{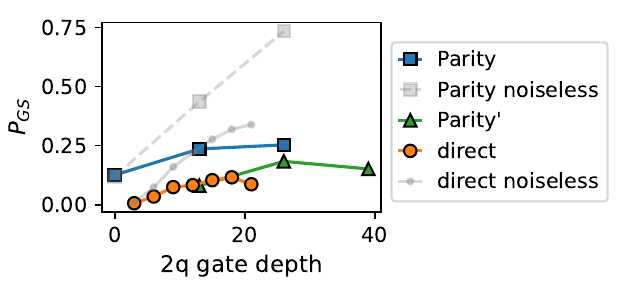}
    \caption{Ground state probability $P_{GS}$ of noiseless simulations (grey) and hardware experiments on \texttt{ibm\_torino} (colored) as a function of the two-qubit gate depth for a heavy-hex graph of $N=21$ nodes. The first marker correspond to $p=1$, the second marker to $p=2$ and so on.
    Best parameters obtained from the classical simulations were used for a single job with $4096$ measurements.
    For {\Parity} QAOA one spanning tree $M=1$ is used for decoding.
    For the green line \textit{Parity'} the first constraint unitary is applied despite $\Omega_1=0$, showing the impact of noise from the additional constraint layer.}\label{fig:Results_HeavyHex_N100}
\end{figure}

\subsection{Hardware-native graph}\label{sec:results:hardware_native}

\subsubsection{Circuits}
As a first example, we consider the hardware-native graph given in Fig.~\ref{fig:HeavyHex_coveringdepth_parameters}(a) and its {\parity} transformation shown in Fig.~\ref{fig:HeavyHex_coveringdepth_parameters}(b).
The {\parity} representation in Fig.~\ref{fig:HeavyHex_coveringdepth_parameters}(b) uses $K = 22$ physical qubits for $N = 21$ 
logical variables and
requires $L= K - N + 1 = 2$ independent constraints. 
Both constraints act on $12$ qubits and correspond to the two cycles in the logical graph that share the edges  $(4, 10)$ and $(10, 16)$. 
The implementation of the entangling gates for both methods is shown in Fig.~\ref{fig:HeavyHex_coveringdepth_parameters}(c). The logical graph uses only two-qubit interactions, which are implemented via $R_{zz}$ gates (top). The two parity constraints each involve $N_{c_l}=12$ qubits and can be implemented using CNOT chains and an $R_{zz}$ gate (bottom). For simplicity, an example with six qubits is shown, but the same mechanism extends to any {$N_{c_l}$-body} constraint: CNOT chains accumulate the required parities on two qubits, an $R_{zz}$ gate applies the phase, and the CNOT chains are reversed. In this setting, the constraints can be represented by paths on the physical layout, rather than the plaquettes used in previous implementations~\cite{bridge_compiler}.

In this simple example, each logical qubit in Fig.~\ref{fig:HeavyHex_coveringdepth_parameters}(a) interacts only with its two (or three) neighbors, whereas
each parity constraint in Fig.~\ref{fig:HeavyHex_coveringdepth_parameters}(b) involves twelve qubits. 
The reported covering depths are ${c_p=2}$ for the
{\parity} representation and ${c_p=10}$ for the logical representation, but the smaller $c_p$ comes at the cost of a larger two-qubit depth per {\parity} QAOA layer. For the compilation considered here, the {\parity} implementation has a two-qubit gate depth of $12$, with CNOT chains grown from both ends and gates from the two constraints scheduled in parallel. For the direct implementation, each QAOA layer has a depth of 3, obtained by scheduling its interactions in three sets of non-overlapping edges.
In addition, the {\parity} implementation requires approximately twice as many two-qubit gates (40 gates) as the direct implementation (22). 

Considering two-qubit gate depth and two-qubit gate count {direct} QAOA requires fewer resources per entangling
QAOA layer, while {\Parity} QAOA has the smaller covering depth. 
To assess how these competing
properties affect optimization performance, we now turn to QAOA simulations.

\subsubsection{Noiseless simulations}
For the given hardware-native graph, we consider the one ``inequivalent" problem that cannot be exactly solved with Clifford {\parity} QAOA circuits.
Equivalent instances are related by the unitary transformations described in  Appendix~\ref{sec:app:unique_problem}. In addition, in an attempt to select problem instances that are challenging for Parity QAOA, we exclude instances that are solvable with Clifford QAOA circuits. Ref.~\cite{Weidinger_NJP2026} showed that, using a specific set of parameters, these instances can be solved exactly with {\parity} QAOA.

Noiseless simulations are performed using Qiskit with $10^4$ measurement shots. We start from
random parameter values, optimize them using COBYLA~\cite{ragonneau_thesis, ragonneau_cobyqa} and take the best run. For Parity QAOA, we use one spanning tree $M=1$ for decoding. Further details are given in Appendix~\ref{sec:app:simulation_details}.

The best parameters found for $p\in\{1, 2, 3\}$ are shown in Fig.~\ref{fig:HeavyHex_coveringdepth_parameters}(d). 
The results for $p=1$ are represented by a single marker at $i=1$. Parameters for $p=2$ and $p=3$ are denoted by two and three markers that are connected via a dashed and solid line, respectively.
For the direct implementation (top panel), they follow the observed trend of the annealing schedule: The $\beta_i$ values decrease and the $|\gamma_i|$ values increase with increasing $i$~\cite{Zhou_Lukin_prx2020, MontanezBarrera_npjQuantum2025}.
The parameters for {\parity} QAOA  are shown in the bottom panel of Fig.~\ref{fig:HeavyHex_coveringdepth_parameters}(d). The best-found first-layer constraint parameters are close to zero, which is consistent with previous works~\cite{Wang_Weidinger_arXiv2026, Wybo_Quantum2024, Weidinger_NJP2026}. Setting $\Omega_1 = 0$ removes the first entangling constraint layer and leaves a product-state preparation
that can be sampled efficiently on a classical computer. Nonzero constraint parameters at later layers can introduce entangling operations.
Similarly to direct QAOA, for $1<i\leq3$ the parameters $\beta_i$  decrease and the parameters $|\gamma_i|$ increase with increasing $i$.
Explicit values of the parameters are listed in Appendix~\ref{sec:app:simulation_details}.

\begin{figure*}
    \centering
    \includegraphics[width=1\linewidth]{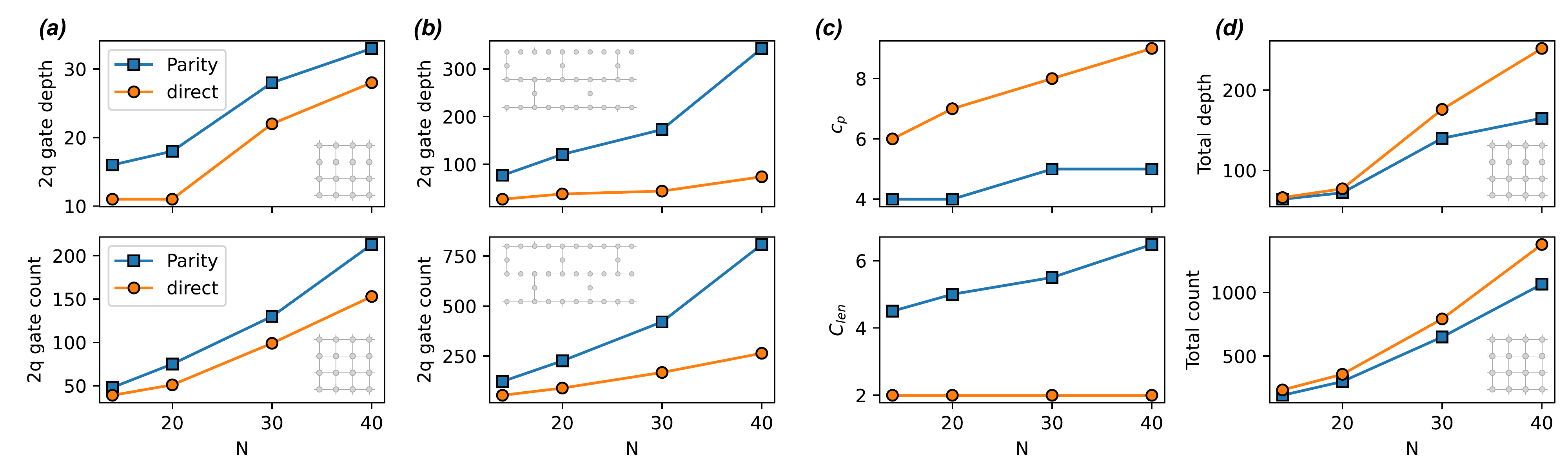}
    \caption{Two-qubit gate depth and two-qubit gate count (with $R_{zz}$ gates) for the first entangling layer of random 3-regular graphs on a square grid (a) and on a heavy-hex topology (b). The top panel in (c) shows the covering depth $c_p$ and the bottom panel shows the number of qubits in the entangling gates (constraints, interactions). Panel (d) shows the total two-qubit gate depth (two-qubit gate depth multiplied by $c_p$) and total two-qubit gate count needed to cover the whole graph on a square grid layout.}
    \label{fig:circuit_specifics}
\end{figure*} 

\subsubsection{Experiments}
We use the best-found parameters to run experiments on IBM hardware, which has a heavy-hex topology. 
The circuits are executed on \texttt{ibm\_torino}, with $4096$ measurements. 
The results are shown in Fig.~\ref{fig:Results_HeavyHex_N100} together with the result of the noiseless simulation results (in grey) for reference. 
We show the probability $P_{GS}$ of sampling a ground-state configuration, as a function of the two-qubit gate depth. This comparison accounts for the
different two-qubit depths per QAOA layer of the two algorithms: $3$ for direct QAOA and $12$ for the Parity implementation. 
The first marker corresponds to $p=1$, the second marker to $p=2$ and so on. 
The label {\Parity}' denotes a run in which the first constraint unitary is retained
despite the parameter $\Omega_1$ being null, illustrating the noise introduced by these additional gates.

The logical graph fits natively on the hardware, the local interactions can be effectively parallelized, leading to a significantly smaller two-qubit depth per QAOA layer than the parity implementation ($3$ vs $12$).
However, Parity QAOA achieves a higher ground-state sampling probability than direct QAOA
in both noiseless simulations and hardware experiments, despite its larger two-qubit depth per QAOA
layer.
The highest ground state probability $P_{GS}$ is reached when $p=3$ for {\parity} and $p=2$ for Parity', which is at the same two-qubit gate depth. The direct implementation performs best at $p=6$ but has, in general, a lower ground state probability $P_{GS}$ than {\parity} QAOA. These results confirm that the observed benefit of the
{\Parity} mapping is not limited to problems with a connectivity mismatch.

Even though the heavy-hex graph is native to the hardware, it profits from the {\parity} transformation. This shows that {\Parity} QAOA has applications beyond its original purpose. 

To extend the analysis to NP-hard problems, the next section investigates the performance of both QAOA methods on random 3-regular graphs.

\subsection{3-regular graphs}\label{sec:3-regular_graphs}

\subsubsection{Circuits}
We generate random 3-regular graphs of sizes ${N\in\{14, 20, 30, 40\}}$ with the Python package NetworkX, rejecting planar graphs and accepting the first non-planar graph for each investigated problem size.

We compile both QAOA implementations for square-grid and heavy-hex connectivity. For direct QAOA, which
requires routing for these non-native graphs,
we use the Qiskit transpiler at optimization level 3 to generate the QAOA circuits, keeping the best of six compilation attempts. 
For {\parity} QAOA, we use ParityOS~\cite{parityos} to generate the QAOA circuits, with a flexible constraint compilation strategy~\cite{bridge_compiler}. 
Both implementations are then transpiled for the heavy-hex and square grid connectivity by using the Qiskit pass manager at optimization level 3 and again retaining the best out of six
attempts. Circuit specifics are presented in Fig.~\ref{fig:circuit_specifics}.

\begin{figure*}
    \centering
    \includegraphics[width=0.9\linewidth]{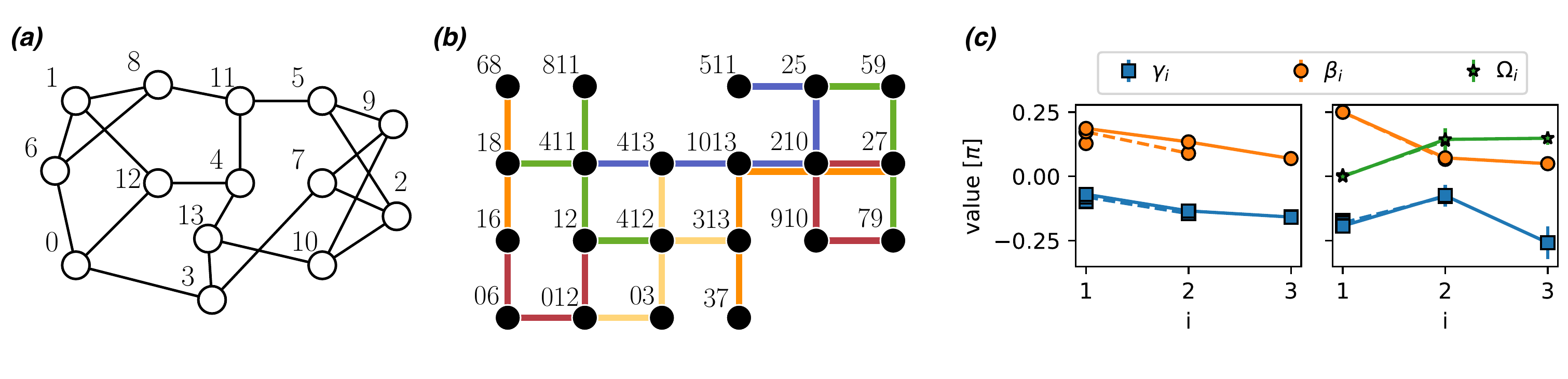}
    \caption{3-regular graph with $N=14$ nodes in its (a) logical and (b) Parity representation. Panel (c) shows the average of the best found parameters for both QAOA variants. Parameters for $p=1$ are presented with a single marker and for $p=2$ and $p=3$ the markers are connect by a dashed and solid, respectively. Error bars denote the standard deviation.}
    \label{fig:results_3regular}
\end{figure*}

For the circuits considered here, the square-grid connectivity (see Fig.~\ref{fig:circuit_specifics}(a)) yields smaller two-qubit gate counts and depths
than the heavy-hex connectivity (see Fig.~\ref{fig:circuit_specifics}(b)). This additional overhead is more pronounced for Parity QAOA than for direct QAOA. For $N = 40$, the gate-count
ratio between the Parity and the direct QAOA implementation increases from $1.39$ for the square grid to $3.06$ for the heavy-hex connectivity, while the depth ratio increases from $1.18$ to $4.64$.

The parity compilers used in this publication are optimized for rectangular grids, which means that the resulting circuits have to be further transpiled to a heavy-hex topology. The long constraints are then difficult to route along this sparse topology. Here, special-purpose parity compilers could be developed to optimize parity compilation with long constraints specifically for heavy-hex grids. However, most quantum computers offer a square grid topology and IBM now provides square-lattice topologies with the Nighthawk generation of QPU as well. To benchmark the algorithms, we use the heavy-hex connectivity of the Heron devices~\cite{IBM_quantum}, which are accessible via IBM's open access plan.

The top panel of Fig.~\ref{fig:circuit_specifics}(c) depicts the covering depth $c_p$ or each graph instance considered in the previous two panels. The covering depth differs only by 2 for $N=14$, but the difference between the direct graph and the Parity hypergraph increases with larger system sizes, because the covering depth $c_p$ for the {\parity} hypergraph increases slower than the one for the logical graph. This is due to the average number of qubits in their entangling gates, shown in the bottom of panel (c). The constraints grow with increasing system size $N$, because the logical graph becomes more sparse and that creates (on average) larger constraints in the {\parity} hypergraph, while for the direct QAOA the metric stays constant ($C_{len}=2$). There are, on average, $4.5$ qubits in one constraint for $N=14$ and $6.48$ qubits for $N=40$. 
The total resources needed to cover the whole graph (multiplying the depth and gate count by $c_p$) on a square grid are visualized in (d), showing that the advantage of the {\parity} implementation increases with increasing $N$. In order to see whether this observed advantage for Parity QAOA, which comes with a qubit overhead of $K=1.5N$, equates to an advantage in QAOA performance, we perform noiseless simulations and hardware experiments.

\subsubsection{Noiseless simulations}
We perform noiseless QAOA simulations on the 3-regular graph of size $N=14$, with Qiskit, using $10^4$ measurement shots. More simulation details can be found in the Appendix~\ref{sec:app:simulation_details}. In general, for 3-regular graphs, the parity graph has $K=1.5N$ physical qubits, which means $K=21$ qubits for $N=14$. The graph and its Parity representation are shown in Fig.~\ref{fig:results_3regular}(a) and (b), respectively.
We randomly generate 12 inequivalent Signed Max-Cut training instances for this graph, excluding instances that are solvable with Clifford QAOA circuits.

\begin{figure*}
    \centering
    \includegraphics[width=1\linewidth]{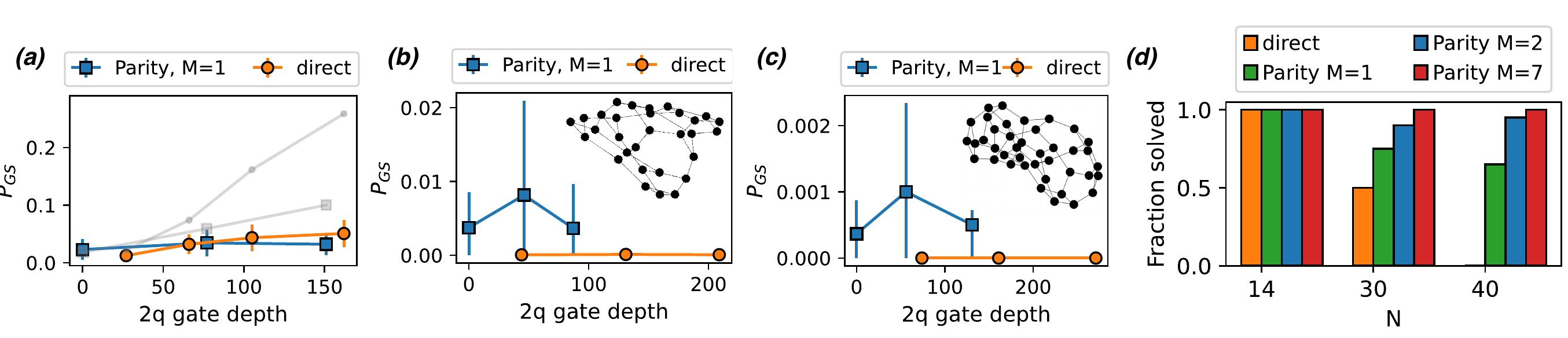}
    \caption{
    Ground state probability $P_{GS}$ obtained from experiments on the \texttt{ibm\_fez} machine for 3-regular graphs of size (a) $N=14$ , (b) $N=30$ and (c) $N=40$. The solid lines denote the average over 12 random instances for (a) and 20 random instances for (b) and (c), error bars denote the standard deviation. 
    The first marker correspond to $p=1$, the second marker to $p=2$ and so on.
    Results from noiseless simulations for $N=14$ are represented by gray lines in (a), square markers for parity and dots for the direct implementation. The simulated graphs for $N>14$ are shown in the inlet.
    Panel (d) shows the fraction of instances where the QAOA algorithm measured the ground state at least once (considering all $p$). One can increase the fraction for {\Parity} QAOA by using more spanning trees $M$ for decoding. At $M=7$ all instances are solved for all $N$.}
    \label{fig:results_3regular_experiment}
\end{figure*} 

 Fig.~\ref{fig:results_3regular}(c) shows the best-found pentameters across the 12 training instances for $p=1,2,3$ QAOA layers.  
For {\parity} QAOA the values for the different layers $p$ are so similar that the difference cannot be seen in the plot. The difference is also small, but visible in the plot, for direct QAOA. 
For Parity QAOA, the parameter $\Omega_1$ is null, $\Omega_1=0$. This is the same behavior we observed in the previous section, when optimizing the heavy-hex graph. The absolute value of $|\Omega_{p>1}|$ and $|\gamma_{p>1}|$ increases with $p$ and the value for $\beta_{p>1}$ decreases, showing that {\parity} QAOA follows an annealing schedule similar to direct QAOA at least for small $p$. The small spread of the QAOA best-found parameters across the different training instances is consistent with the emergence of a parameter concentration phenomena for both direct and Parity QAOA~\cite{Wurtz_PRA2021}. 
We use the mean parameters in the subsequent hardware experiments, including
the larger instances.

Minimizing the mean decoded energy in Eq.~\eqref{eq:C_parity_min} does not always lead to maximizing the ground-state probability~\cite{Shaydulin_SciAdv2024, Barkoutsos_Quantum2020}. 
In particular, Parity QAOA's optimal parameters can concentrate on the ground state of the $1$-local Hamiltonian $\tilde{H}_z $ (Eq.~\eqref{eq:local_field_hamiltonian}). Its decoded energy can be favorable even when it does not produce an
optimal logical configuration. To discourage concentration on this state, we use a modified objective that
penalizes its contribution. In Appendix~\ref{sec:app:3regular_parameter_objective}, we provide the definition of the modified objective and compare this approach with an explicit maximization of $P_{GS}$.

\subsubsection{Experiments}
We use the previously compiled circuits and the average of the best-found parameters of the previous simulations for a single run with 4096 measurements on the machine \texttt{ibm\_fez}. 
First, we compare the experimental results of the graph with $N=14$ nodes as a function of the two-qubit gate depth (considering heavy-hex connectivity) to the noiseless ones. They are shown in Fig.~\ref{fig:results_3regular}(a). 
Parity QAOA has a slight increase in ground state probability $P_{GS}$ going from $p=1$ to $p=2$, but is unable to show significant improvement due to noise. Direct QAOA can achieve a similar ground state probability $P_{GS}$.

For the larger instances studied here, hardware noise prevents a second layer of the full {\parity} QAOA circuit from improving the algorithm's success probability. We therefore truncate the constraint circuits and implement only selected subset of constraints in each layer, reducing the two-qubit gate count and the noise exposure. Denoting the corresponding constraint index set for the $i$-th layer by $S_i$, the implemented constraint unitaries are
$
\tilde{U}_{\squaredots}^{(i)}(\Omega_i)
=
\prod_{l\in S_i}\eeuler^{-i\Omega_i\tilde{H}_{\squaredots,l}}$.
Specifically, for $N=30$ and $N=40$, we chose $S_1$ and $S_2$ that satisfy $|S_1| = |S_2| \approx \lfloor L/4 \rfloor $, such that each subset contains approximately a quarter of the full constraint set.
Together, these subsets contain approximately half of the full constraint set while the remaining constraints are omitted. We do not apply a similar truncation to direct QAOA because adding a second layer to its full circuit reduces the mean cost function despite hardware noise.

The results for the larger 3-regular graphs with $N=30$ and $N=40$ are shown in Fig.~\ref{fig:results_3regular_experiment}(b) and (c), respectively. The inlet shows the selected graphs. We consider $20$ random instances of the Signed Max-Cut problem for each graph.

With the circuit truncation {\parity} QAOA is able to improve the average ground state probability $P_{GS}$ from $p=1$ to $p=2$ for all considered system sizes. 
For $N>14$, the ground-state probability for direct QAOA is indistinguishable from zero in Fig.~\ref{fig:results_3regular_experiment}(b) and (c).
Figure~\ref{fig:results_3regular_experiment} (d) shows the fraction of solved instances, illustrating that {\parity} QAOA is able solve every single instance when using $M=7$ spanning trees for decoding. Direct QAOA did not sample a solution to any of the instances for $N=40$, but could sample a solution in half of the problem instances of size $N=30$. Compared to direct QAOA the decline in success probability with increasing size is less pronounced for the {\parity} QAOA 
pipeline over the tested instances.

Note, the Parity transformation offers the possibility for decoding as a result of its qubit overhead. For 3-regular graphs, this overhead scales as $1.5N$, which is significantly smaller than the $N^2$ overhead of the Parity transformation applied to a complete graph.

{\Parity} QAOA effectively implements a warm-start by selecting $\Omega_1=0$. This changes the starting state from an equal superposition of all states to a different product state. In Fig.~\ref{fig:results_3regular_experiment} the first marker in the panels (a)-(c), show that this warm-start returns a non-zero ground state probability for small system sizes $N\leq40$ but naturally raises the question of how the quality of the starting state scales with increasing problem size $N$.
In Appendix~\ref{sec:app:starting_state}, we evaluate the quality of this starting state for system sizes up to $N=500$. The results suggest that its quality remains stable for larger instances.

\section{Conclusion and Outlook}\label{sec:conclusion}
In this work, we have investigated the performance of QAOA on sparse graphs using noiseless numerical simulations and experiments on IBM Heron quantum processors. Specifically, we compared two implementations: direct QAOA and Parity QAOA.

Building on previous work, we relaxed the restriction to short constraints for {\parity} QAOA by allowing constraints that act on more than four qubits. These higher-order constraints cover larger portions of the graph, but come at the cost of increased two-qubit circuit depth per layer. Despite this tradeoff, on the hardware-native instance studied here, {\parity} QAOA sampled optimal solutions with a higher probability than direct QAOA. This extends the original purpose of the {\parity} Architecture beyond addressing the connectivity mismatch between the problem graph and the hardware. 

For 3-regular graphs, the experiments demonstrated a sampling advantage of {\parity} QAOA over direct QAOA, which becomes more pronounced as the problem size increases. In particular, {\parity} QAOA sampled an optimal solution for
every studied $40$-vertex $3$-regular graph instance, whereas direct QAOA sampled none within the same measurement budget.

Interestingly, the best-found Parity QAOA parameters have first-layer constraint angles close to zero. Setting this angle to zero produces  a product state that effectively serves as a warm start and whose measurement distribution can be sampled efficiently on a classical computer. This
baseline already yields useful decoded solutions, making further improvements with noisy entangling layers challenging.
 However, by truncating the circuit and omitting some of the constraints,  we observed improvements over the product-state baseline for the tested larger instances.  This highlights the challenge of achieving a clear advantage over classical algorithms on noisy quantum processors, where the benefit of additional entangling layers must outweigh the noise they introduce. Further advances in quantum hardware, circuit design, and error mitigation may help address this tradeoff.

The small spread of optimized parameters across the 14-vertex training instances supports their use as transferable parameters in the experiments. However, establishing parameter concentration with system size for {\parity} QAOA and understanding how the parameters depend on circuit truncation requires further study. 

Overall, our results identify the optimization problems on sparse graphs as a promising setting for Parity QAOA with higher-order constraints. The observed improvements are consistent with a beneficial role of the shorter covering depth associated with these constraints.

\vspace{5mm}
Parity QAOA circuits considered in this work are available online \cite{GitlabTutorial}.

\section*{Acknowledgements}
The authors thank M. Fellner, R. Banks and R. Wang for useful discussions. This study was supported by NextGenerationEU via FFG and Quantum Austria (FFG Project No. FO999896208).
This research was funded in part by the Austrian Science Fund (FWF) under Grant-DOI 10.55776/F71 and Grant-DOI 10.55776/Y1067.
This project was funded within the QuantERA II Programme that has received funding from the European Union’s Horizon 2020 research and innovation programme under Grant Agreement No. 101017733. 
This publication has received funding under Horizon Europe programme HORIZON-CL4-2022-QUANTUM-02-SGA via the project 101113690 (PASQuanS2.1)
This study was supported by the Federal Ministry for Economic Affairs and Climate Action through project QuaST.
For the purpose of open access, the author has applied a CC BY public copyright license to any Author Accepted Manuscript version arising from this submission.
The authors acknowledge the use of IBM Quantum services for this work. The views expressed are those of the authors, and do not reflect the official policy or position of IBM or the IBM Quantum team.

\appendix

\section{Equivalent problem instances}\label{sec:app:unique_problem}
If the ground states of the local field $H_z(J_{ij})$ and $H_z(J'_{ij})$ violate the same constraints, they behave the same under {\parity} QAOA, meaning that some parameters ($\beta, \Omega, \gamma$) return the same approximation ratio or ground state probability. Given a unitary transformation W that transforms from one problem to the other $J_{ij} \to J'_{ij}$ then the following equation only holds if $U_{\squaredots}(\Omega)$ is invariant under the transformation ($U_x(\beta)$ is invariant in any case),
\begin{equation}
    U_z(\gamma, J_{ij} )U_{\squaredots}(\Omega) = W U_z(\gamma, J'_{ij}) W^{-1}U_{\squaredots}(\Omega).
\end{equation}

Therefore, there is one unique problem instance that represents several other problems that violate the same constraints. If problem instances are inequivalent in Parity QAOA, they are also for direct QAOA.

\begin{table*}
\begin{tabular}{ c | c |c| c} 
 p/i & 1 & 2 & 3 \\
 \hline
 1 &  -0.210, 0.015, 0.25 &  & \\
2 & -0.215 -0.001  0.25 & -0.012   0.264  0.032 &\\
3 & -0.201 -0.002 0.25 & -0.041 0.210 0.054 & -0.203 0.193 0.028\\
 \hline
1 & -0.173 (0.007) 0.004 (0.002) 0.25  &  & \\
2 & -0.181 (0.007) 0.001 (0.003) 0.25& -0.079 (0.029) 0.139 (0.023) 0.067 (0.007) & \\
3 & -0.193 (0.011) -0.001 (0.005) 0.25 & -0.075 (0.043) 0.144 (0.043) 0.072 (0.008) & -0.259 (0.065) 0.149 (0.027) 0.050 (0.009)\\
\hline
\end{tabular}
\caption{Best parameters found $(\gamma_i^*, ,\Omega_i^*, \beta_i^*)$ for {\parity} QAOA for the heavy-hex graph $M=1$, $N=21$, (first three rows) and the 3-regular graph, $N=14$ and $M=2$ (last three rows). Values for the 3-regular graphs are the average over 12 instances, the standard deviation is given in brackets . Units in $\pi$. }\label{tab:app:best_parameter_parity}
\end{table*}

\begin{table*}
\begin{tabular}{ c | c | c | c | c} 
 p/i & 1 & 2 & 3 & 4 \\
 \hline
 1 &  -0.115  0.123 &  & & \\
 2 &-0.101  0.190 & -0.184  0.104 & &\\
3 & -0.086 0.196 & -0.162 0.159 & -0.194 0.073 &\\
4 & -0.169  0.427 & 0.091  0.271  &-0.174  0.135 &-0.192   0.076\\
 \hline
1 &  -0.098 (0.002)  0.128 (0.055) &  & & \\
2 & -0.080 (0.004) 0.173 (0.009) & -0.143 (0.006) 0.091 (0.075) & &\\
3 & -0.072 (0.004) 0.181 (0.011) & -0.136 (0.012) 0.129 (0.019) & -0.160 (0.019) 0.065 (0.011) & \\
4 & -0.061 (0.006) 0.191 (0.01) & -0.116 (0.010) 0.142 (0.014) & -0.141 (0.015) 0.106 (0.018) & -0.166 (0.023) 0.055 (0.008) \\
 \hline
\end{tabular}
\caption{Best parameters found $(\gamma_i^*, \beta_i^*)$ for direct QAOA for the heavy-hex graph $M=1$, $N=21$, (first four rows) and the 3-regular graph, $N=14$ and $M=2$ (last our rows). Values for the 3-regular graphs are the average over 12 instances, the standard deviation is given in brackets . Units in $\pi$.}\label{tab:app:best_parameter_direct}
\end{table*}

\section{Simulation details and parameter results}\label{sec:app:simulation_details}
All QAOA simulations are done with Qiskit. The parameter optimization works in the following way: 
We propose several initial parameters and optimize them further with COBYLA (from SciPy)~\cite{ragonneau_thesis, ragonneau_cobyqa}. Out of the several optimization runs we take the best one to calculate statistics. 
For the heavy-hex graph we use $40$, $80$ and $320(p-1)$ random start parameters $\bfbeta, \bfgamma$ (and $\bfOmega$ for {\parity}) for for $p=1$, $p=2$ and $p\geq3$, respectively.
For the 3-regular graphs, we use $20/160/480/800$ optimization runs for $p=1/2/3/4$.

The lowest (and the highest) energy for bigger graphs cannot be calculated with brute force. For the heavy-hex graph, which is planar, we can use Minimum weight perfect matching (MWPM) to get these energies. For the 3-regular graphs, which are non-planar, we get them with the Gurobi solver~\cite{gurobi} (MILP).

Explicit values of the best parameters found by optimization are given in Tab.~\ref{tab:app:best_parameter_parity} and Tab.~\ref{tab:app:best_parameter_direct} for {\parity} and direct QAOA, respectively. 
The obtained parameters for direct QAOA are consistent with those presented in Ref.~\cite{Wurtz_PRA2021} within error bars (and considering a factor of 2 in $\gamma$ due to Hamiltonian definition).

\begin{figure}
    \centering
    \includegraphics[width=0.85\columnwidth]{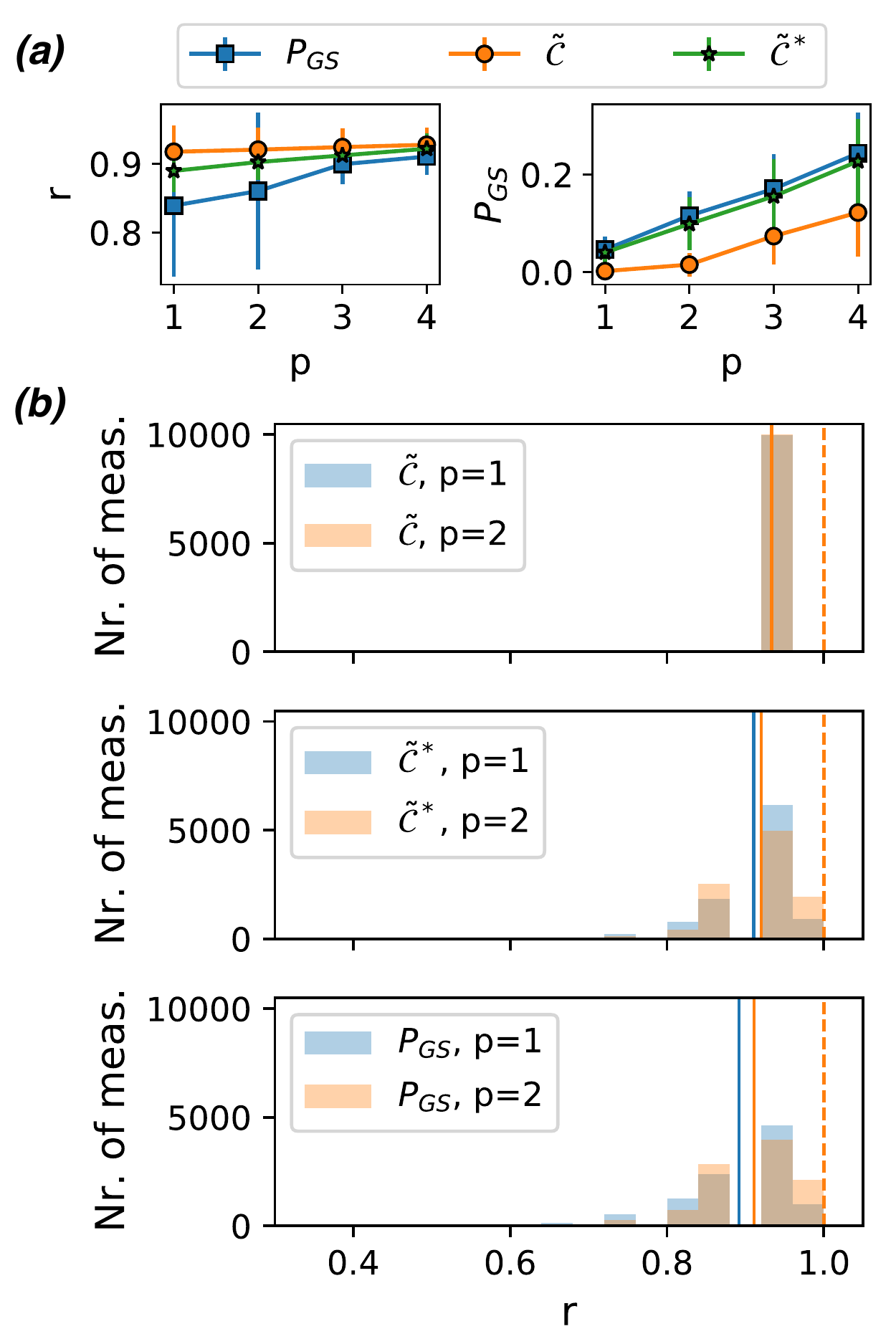}
    \caption{Comparison of Parity QAOA with different cost functions. The three different objectives are the energy $\tilde{\mathcal{C}}(\gamma, \beta, \Omega)$, the ground state probability $P_{GS}(\gamma, \beta, \Omega)$ and the adapted energy $\tilde{\mathcal{C}}(\gamma, \beta, \Omega)^*$ (where the energy of the local field ground state gets an energy penalty by 20\%.) Panel (a) shows the obtained approximation ratio $r$ (left) and ground state probability $P_{GS}$ (left) from noiseless simulations in dependence of QAOA layer $p$, $M=2$, averaged over 12 instances.  
    Error bars denote the standard deviation. Panel (b) shows histograms of measurement results of a single instance for the three different objective functions. Parity QAOA with the default objective can get stuck by measuring a single state (top). A change in the objective function can avoid this (middle and bottom).}
    \label{fig:app:Optimization_3reg_parity_comparison}
\end{figure}


\begin{figure}
    \centering
    \includegraphics[width=0.85\columnwidth]{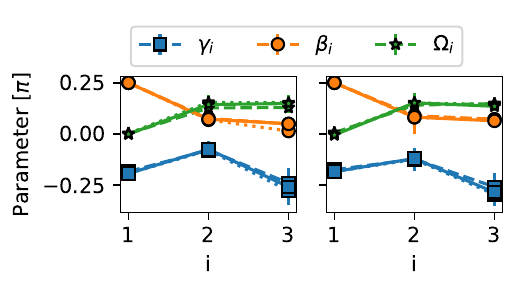}
    \caption{Best found parameter for $p=3$ for Parity QAOA optimized with $\tilde{\mathcal{C}}(\gamma, \beta, \Omega)^*$ (left) and $P_{GS}(\gamma, \beta, \Omega)$ (right). Best parameters of QAOA simulations with $M\in\{1, 2, 3\}$ are shown, with lines dashed, solid  and dotted, respectively.}
    \label{fig:app:ParameterComparison_Parity}
\end{figure}

\section{Parameter for Parity QAOA and different objective functions}\label{sec:app:3regular_parameter_objective}
In this section, we compare three different objective functions for {\parity} QAOA. We consider the original objective function $\tilde{\mathcal{C}}(\gamma, \beta, \Omega)$, Eq.~\eqref{eq:C_parity_min}, 
an adapted objective function $\tilde{\mathcal{C}}(\gamma, \beta, \Omega)^*$, (where the ground state of the local field $\tilde H_z$ gets an $20\%$ energy penalty) 
and the ground state probability $P_{GS}(\gamma, \beta, \Omega)$. 
Another possible objective function is the CVaR cost function, which focuses only on a subset of low-energy samples and not on the whole distribution, but its testing is beyond the scope of this project~\cite{Barkoutsos_Quantum2020}.

As mentioned in the main text, when using the original cost function $\tilde{\mathcal{C}}(\gamma, \beta, \Omega)$ {\parity} QAOA tends to fall into a trivial minimum, where one only measures the ground state of the local field $\tilde H_z$. This state can be reached with the classical Clifford parameters ($\gamma_1, \Omega_1, \beta_1$)=($-0.25\pi, 0, 0.25\pi$). As this classical state gives a reasonable energy (compared to a superposition), {\parity} QAOA falls into this trivial minimum for half of the considered instances, even with $p>1$. 
By adapting the cost function so that this single state experiences an energy penalty, one can avoid this problem. 
To be more precise, when the variational state is a linear combination of eigenstates of the total Hamiltonian $\tilde{H}_P$
\begin{equation}
\ket{\tilde{\Psi}(\bfbeta,\bfOmega,\bfgamma)} = c_0\ket{\tilde \phi_0} + c_1\ket{\tilde \phi_1} + ... + c_{2^k}\ket{\tilde \phi_{2^k}},
\end{equation}
and define $\ket{\tilde \phi_0}=\ket{\tilde \psi_{H_z, 0}}$ as the ground state of the local field Hamiltonian $\tilde{H}_z $ (Eq.~\eqref{eq:local_field_hamiltonian}) and the total cost function $\tilde{\calC}(\bfbeta,\bfOmega,\bfgamma)=\tilde{\calC}_0 + \tilde{\calC}_1 + ... + \tilde{\calC}_{2^K}$ , and $\tilde{\calC}_0=\tilde{\calC}_{H_z, 0}$, then the modified cost for this specific state is
\begin{align}
     \tilde{\calC}_0^*=1.2\tilde{\calC}_0
     =1.2 \ \underset{t \in T}{\min} &  \left( H_P \,D_{t}[
     \ket{\tilde \psi_{H_z, 0}}]\right)\label{eq:adapt_energy_local_field_gs}.
\end{align}

The adapted cost function is $\tilde{\calC}(\bfbeta,\bfOmega,\bfgamma)^*=\tilde{\calC}_0^* + \tilde{\calC}_1 + ... + \tilde{\calC}_{2^K}$.

The results of QAOA simulations are shown in Fig.~\ref{fig:app:Optimization_3reg_parity_comparison} (a).
Here, we also show results for the normalized \textit{approximation ratio} $r$ which is defined as
\begin{equation}\label{eq:app:approx_ratio}
r = \frac{C_\rmmax - \tilde{\calC}(\bfbeta,\bfOmega,\bfgamma)}{C_\rmmax -C_\rmmin},
\end{equation}
where $C_\rmmax$ and $C_\rmmin$ are the maximum and minimum value of the objective function. This ratio quantifies how close $\tilde{\calC}(\bfbeta,\bfOmega,\bfgamma)$ is to the optimal value $C_\rmmin$, where ${r = 1}$ corresponds to the optimal solution.

We observe that the approximation ratio $r$ for Parity QAOA with the usual objective $\tilde{\mathcal{C}}(\gamma, \beta, \Omega)$ does not improve significantly with increasing $p$, but it achieves the highest approximation ratio $r$ of all three methods. However, its achieved ground state probability is the lowest. The other two methods return a similar ground state probability, but optimizing with $\tilde{\mathcal{C}}(\gamma, \beta, \Omega)^*$ leads to a better approximation ratio. Thus, we show the results of $\tilde{\mathcal{C}}(\gamma, \beta, \Omega)^*$ in the main text. However, we do not conclude that this method works well for more layers $p$ or other problem graphs.

Figure~\ref{fig:app:Optimization_3reg_parity_comparison}(b) shows histograms of the results of the measurements of a single instance for $p=1$ and $p=2$ for all three optimization methods. The solid line denotes the average energy and the dashed line the best measured energy of the distribution. 
The top panel shows that using the original objective function $\tilde{\mathcal{C}}(\gamma, \beta, \Omega)$ can lead to a trivial minimum, that is, only measuring the local field ground state. Hence, we observe only a single peak. A second layer, $p=2$ cannot improve on it. This is in line with observations that QAOA with a classical starting state gets stuck~\cite{Cain_arXiv2023}.
Adapting the objective function $\tilde{\mathcal{C}}(\gamma, \beta, \Omega)^*$ avoids getting stuck in this trivial minimum, as shown in the middle panel. Similar results can be achieved when using the ground state probability $P_{GS}(\gamma, \beta, \Omega)$ as an objective function (bottom panel). 


The parameters obtained for $p=3$ when using the adapted objective function $\tilde{\mathcal{C}}(\gamma, \beta, \Omega)^*$ and the ground state probability $P_{GS}$ are presented in Fig.~\ref{fig:app:ParameterComparison_Parity}. This also includes results when using different numbers of spanning trees $M$ for decoding. The Figure shows that all parameters are similar.

\section{Scaling of the starting state}\label{sec:app:starting_state}
In this work, we observe that Parity QAOA does not apply the first constraint layer as the best parameter found for it is $\Omega_1=0$. Previous works observed the same behavior~\cite{Weidinger_NJP2026, Wybo_Quantum2024, Wang_Weidinger_arXiv2026}. 

The single qubit rotations in Parity QAOA's first layer change the starting state from an equal superposition of all states to a different state. Hence, Parity QAOA automatically performs a warm-start protocol. 
The new starting state can be generated using a classical algorithm and we use this to assess the performance of this starting state for random 3-regular graphs with system sizes up to $N=500$. Here, the lowest and highest energies are obtained with the gurobi solver~\cite{gurobi}. This exact algorithm scales exponentially (about $0.02$ seconds for $N=40$ and $189$ seconds for $N=500$).

The two parameters that define single qubit rotations are $\gamma_1$ and $\beta_1$. Due to symmetry, the latter can be fixed to $\beta_1=0.25\pi$, see Appendix~\ref{sec:app:symmetries}. 
When $\gamma_1=0$, one is still in the initial state, every qubit is in an equal superposition of up and down. Changing $\gamma_1$ changes this probability. When $\gamma_1=-0.25\pi$, a Clifford parameter, the probability of measuring the qubits aligned to their local field is $P_{align}=1$, which means that one measures a classical state, specifically the ground state of the local field Hamiltonian $\tilde H_z$. Any other value $-0.25\pi<\gamma_1<0$ changes this probability $0.5<P_{align}<1$. 

One can sample states for a given $P_{align}$ and a given $J$ matrix. The qubit $k$ has a chance $P_{align}$ to align with its local field $J_k$. 
To create a `measured state', one has to repeat this for every qubit, namely $K=1.5N$ times. When repeating this 4096 times one gets a distribution of `measured' states, just as in an experiment. Then, one can decode each measured state with $M$ spanning trees. Hence, this classical algorithm scales $\mathcal{O}(NM)$.

\begin{figure}
    \includegraphics[width=1\columnwidth]{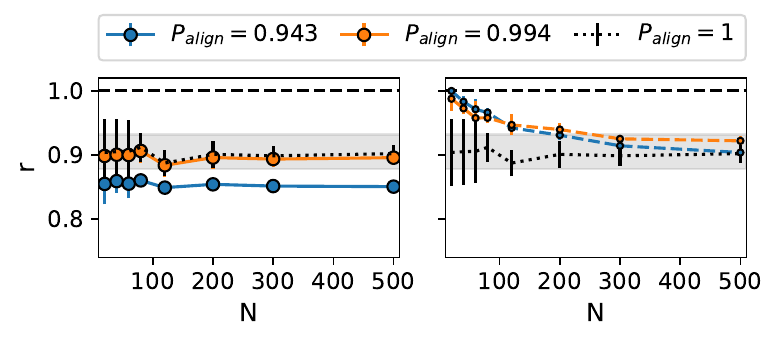}
    \caption{Approximation ratio $r$ of possible starting starts for Parity QAOA with different $P_{align}$ in dependence of system size $N$, $M=1$. 
    Lines correspond to the average of 5 problem instances, error bars are the standard deviation. The left plot in each figure show the average $r$ of a distribution, while the right plot shows the best $r$ of a measured distribution. Grey area markes the GW lower bound of 0.87856 of general graphs and 0.932 for Max-Cut on 3-regular graphs}
    \label{fig:r_vs_N_SQRotations}
\end{figure}

The results of this classical algorithm for 10 different instances $J$ and at each $N$ are shown in Fig.~\ref{fig:r_vs_N_SQRotations} for $M=1$. The plot on the left shows the approximation ratio as defined in Eq.~\eqref{eq:app:approx_ratio} (solid lines), which means that we consider the expectation value $\tilde{\mathcal{C}}(\gamma, \beta, \Omega)$ of the obtained distribution. In the right plot, we see the best measured approximation ratio $r$ (dashed lines) of the obtained distribution. This approximation ratio considers the objective value of a single state (string) $C(\boldsymbol{s})$. The approximation ratio $r$ of the ground state to the local field is related to $P_{align}=1$ and is marked with a black dotted line. Here, the two approximation ratios (left and right plots) coincide. The average approximation ratio $r$ for a distribution $P_{align}<1$ is always lower than that of the classical state $P_{align}=1$ but the best measured approximation ratio for a distribution is better.

The state, created by this simple algorithm, achieves a high approximation ratio, for example, the best measured approximation ratio is $r>0.91$ for every $N$. 
The GW-Algorithm for Max-Cut guaranties an approximation ratio of $r=0.87856$ for general graphs~\cite{GoemansWilliamson_ACM1995} and $r=0.932$ for Max-Cut on 3-regular graphs~\cite{Halperin_JoA2004}. We marked this interval with a gray area in the Figure as a reference.

The results suggest that this starting state is stable for large systems $N$ and that the behavior of leaving out the first constraint layer will probably not change. More {\parity} QAOA layers can increase the approximation ratio, improving the classical lower bound of the warm-start protocol given that one can deal with the introduced noise of the constraint unitary.

\section{Symmetries in Parity QAOA}\label{sec:app:symmetries}
We examine symmetries in Parity QAOA considering that $J_{ij}\in \{-1, 1\}$. 
The unitaries are 
\begin{equation}
    U_x(\beta) = e^{-i\beta \sum_i^K \sigma_i^x},
\end{equation}
\begin{equation}
    U_z(\gamma) = e^{-i\gamma \sum_i^K \sigma_i^z},
\end{equation}
and
\begin{equation}
    U_{\squaredots}(\Omega) = e^{-i\Omega \sum_l^{N_C} \sigma_{l1}^z\sigma_{l2}^z...\sigma_{lk}^z},
\end{equation}
where $N_C$ is the number of parity constraints and $k$ is the number of qubits involved in a constraint. 
Each parameter is periodic in $\pi$, they can be restricted to $\beta, \gamma, \Omega \in(-\pi/2, \pi/2]$. 
This can be seen with (using $\beta$ as an example)
\begin{align}
    U_x(\beta+\pi) &=  \prod_i^K e^{-i(\beta+\pi) \sigma_i^x} \\
    &= \prod_i^K (\cos(\beta+\pi)\mathbb{I}  + i\sin(\beta+\pi)\sigma_i^z) \\
    &= (-1)^N e^{-i\beta \sum_i^K \sigma_i^x} = (-1)^N U_x(\beta),
\end{align}

using the fact that $\cos(\pi+\alpha) = -\cos(\alpha)$ (same for $\sin$). The factor $(-1)^N$ can be neglected as it disappears when evaluating the expectation value. The same proof holds for $\gamma$ and when rewriting the constraint unitary and substituting $S_l= \sigma_{l1}^z\sigma_{l2}^z...\sigma_{lk}^z$ we have

\begin{align}
    U_{\squaredots}(\Omega+\pi) &= \prod_l^{N_C} e^{-i(\Omega+\pi) S_l} \\
    &= e^{-i(\Omega+\pi) S_1}...e^{-i(\Omega+\pi) S_{N_C}},
\end{align}

where, each term can be handled as the previous equation, resulting in a factor of $-1$ if the number of Constraints $N_C$ is odd and $1$ if $N_C$ is even. Again, this factor does not change the expectation value and we can also restrict $\Omega\in(-\pi/2, \pi/2]$.

Considering non-trivial symmetries., we can show that if $\beta_p\to-\beta_p$ then we can compensate with $\gamma_p\to\gamma_p\pm\frac{\pi}{2}$. When rewriting the local field unitary, we see that
\begin{align}
    U_z(\gamma+\frac{\pi}{2}) &= e^{-i\gamma \sum_i^K \sigma_i^z}e^{-i\frac{\pi}{2} \sum_i^K \sigma_i^z} \\
    &= U_z(\gamma) \prod_i^K (\cos(\frac{\pi}{2})\mathbb{I} + i\sin(\frac{\pi}{2})\sigma_i^z) \\
    &= U_z(\gamma) \prod_i^K i \sigma_i^z,
\end{align}

because $\cos(\frac{\pi}{2})=0$ and $\sin(\frac{\pi}{2})=1$. The shift in $\frac{\pi}{2}$ results in a product of $\sigma_z$ over all qubits. Looking at the expectation value the unitaries with the adjusted  $\gamma_p$ and $\beta_p$ are next to the Cost Hamiltonian: 

\begin{equation*}
    U_z(-\gamma) \prod_i^K \sigma_i^z U_x(\beta) (H_z + H_c) U_x(-\beta) \prod_i^K \sigma_i^z U_z(\gamma).
\end{equation*}

Next to the Hamiltonian we can insert the identity and use $\mathbb{I}=\prod_i^K \sigma_i^z\prod_i^K \sigma_i^z$. Then, we drag one product to the other side of the Hamiltonian. Now the unitary $U_x(-\beta)$ is flanked by two $\prod_i^K \sigma_i^z$ and one can show that

\begin{align}
    \sigma_z e^{i \beta \sigma_x}\sigma_z = e^{-i\beta \sigma_x},
\end{align}
because of
\begin{align}
 \prod_i^K \sigma_i^z e^{i\beta \sum\sigma_i^x} \prod_i^K \sigma_i^z &=
  \prod_i^K \sigma_i^z \prod_i \sum_{n=0}^\infty \Big(\frac{i\beta \sigma_i^x}{n!}\Big)^n  \prod_i^K \sigma_i^z.
\end{align}
For even n $\sigma^z \mathbb{I} \sigma^z = \mathbb{I}$ and for odd n  $\sigma^z \sigma^x \sigma^z = -\sigma^x$. With this $\prod_i\sigma_i^z U_x(-\beta)\prod_i\sigma_i^z = U_x(\beta)$ and one can restrict $\beta_p\in[0, \frac{\pi}{2}]$. 

We can also restrict the other $\beta_i$ by using the same trick. For $\beta_p$ we had to insert a $\mathbb{I}$ next to the Hamiltonian to get another $\prod_i^K \sigma_i^z$. For $1<i<p$ when we change $\beta_i \to -\beta_i$ we have to adjust $\gamma_i\to\gamma_i\pm\frac{\pi}{2}$ and $\gamma_{i+1}\to\gamma_{i+1}\pm\frac{\pi}{2}$. Like this $\gamma_{i+1}$ gives another $\prod_i^K \sigma_i^z$ term and one can use the trick as before. So now $\beta\in[0, \frac{\pi}{2}]$.

\begin{figure}
    \centering
    \includegraphics[width=\columnwidth]{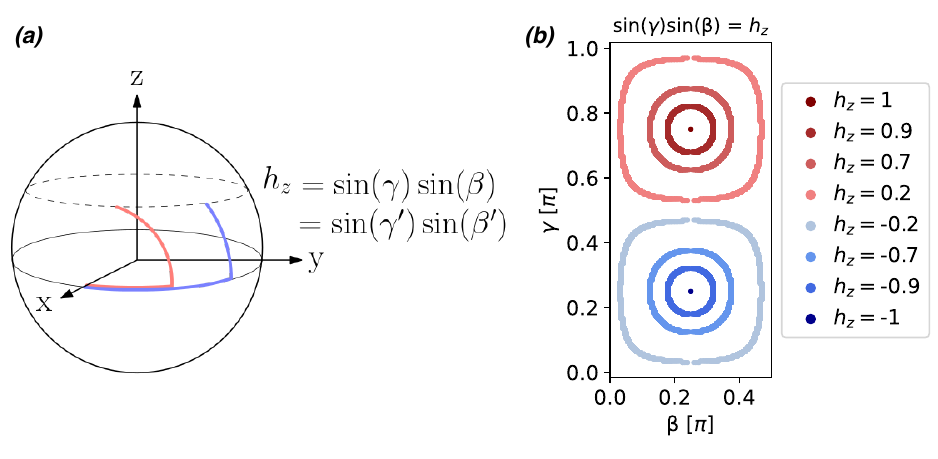}
    \caption{Symmetries for parity QAOA at $p=1$ when $\Omega=0$. The Blochsphere in (a) shows two different rotation sequences $(\gamma, \beta)$ and $(\gamma', \beta')$ (red and blue) where the endpoints are located at the same height $h_z$ on the Blochsphere. The lines in figure (b) mark the combinations of $(\gamma, \beta)$ that end up at a specific height $h_z$. For both plots we assume $J=1$.}
    \label{fig:app:Symmetry_p1}
\end{figure}

In the simulations for the heavyhex and 3-regular graphs we observe that $\Omega_1\approx0$. This induces a symmetry as shown in Fig.~\ref{fig:app:Symmetry_p1}. First, consider $p=1$. There are only single qubit rotations and one can visualize the rotations on the Blochsphere, as shown in Fig.~\ref{fig:app:Symmetry_p1} (a). Here, two rotations sequences are shown: $(\gamma, \beta)$ and $(\gamma', \beta')$ (red and blue lines), considering $J=1$. The endpoints of those two sequences are not the same, but located at the same height $h_z$ on the Blochsphere, making them indistinguishable at a measurement. This means that there are continuous combinations of $(\gamma, \beta)$ that lead to the same measurement outcome. Those combinations, which lead to specific heights $h_z$ in the Blochsphere, are shown in Fig.~\ref{fig:app:Symmetry_p1} (b). 
This symmetry also exists for $p>1$. The two end points in the Blochsphere, Fig.~\ref{fig:app:Symmetry_p1} (a), are separated by an angle $\Delta\gamma$. This difference can be compensated for by adding  $\Delta\gamma$ to the next layer, $\gamma_2$.
Therefore, one can fix $\beta_1=0.25\pi$ and optimize the remaining parameters, assuming that $\Omega_1=0$. Like this, the first layer that was dependent on three parameters now depends only on one parameter: $\gamma_1$.

\end{document}